\documentclass[a4paper,twocolumn]{revtex4}

\usepackage{graphicx,natbib}
\usepackage{mathrsfs}
\usepackage{times}
\usepackage{amsmath}

\newcommand{\bK}{ \mbox{\boldmath$K$} }

\newcommand{\bk}{ \mbox{\boldmath$k$} }

\newcommand{\filter}{ {\cal F} }

\newcommand{\Cref}{ {C_{\rm ref}} }
\newcommand{\ii}{ {{\rm i}} }
\newcommand{\id}{ {{\rm d}} }
\newcommand{\bu}{ \mbox{\boldmath$u$} }

\newcommand{\by}{ \mbox{\boldmath$y$} }
\newcommand{\bDelta}{ \mbox{\boldmath$\Delta$} }
\newcommand{\los}{ \mbox{\boldmath${\ell}$} }
\newcommand{\1}{{\mbox{\boldmath$x_1$}}}
\newcommand{\2}{{\mbox{\boldmath$x_2$}}}

\newcommand{\bx}{{\mbox{\boldmath$x$}}}

\newcommand{\sbx}{{\mbox{\scriptsize\boldmath$x$}}}
\newcommand{\sbk}{{\mbox{\scriptsize\boldmath$k$}}}
\newcommand{\sbDelta}{{\mbox{\scriptsize\boldmath$\Delta$}}}

\newcommand{\zhat}{ \mbox{\boldmath$\hat{z}$} }
\newcommand{\bbv}{ \mbox{\boldmath$v$} }

\newcommand{\bnabla}{ \mbox{\boldmath$\nabla$} }

\begin{document}

\title{F-mode sensitivity kernels for flows}

\author{Jason Jackiewicz}
\author{Laurent Gizon}
\affiliation{Max-Planck-Institut f\"{u}r Sonnensystemforschung, 37191 Katlenburg-Lindau, Germany}
\author{Aaron C. Birch}
\affiliation{NWRA, CoRA Division, 3380 Mitchell Lane, Boulder, CO 80301, USA}

\keywords{Sun; Time-distance helioseismology; f modes, travel-time sensitivity kernels}

\begin{abstract}
We compute f-mode travel-time sensitivity kernels for flows.  Using a
two-dimensional model, the scattered wavefield is calculated
in the first Born approximation. We test the correctness
of the kernels by comparing an exact solution (constant flow), a solution linearized in the flow, and the total integral of the kernel. In practice,  the linear approximation is acceptable for flows as large as about $400$~m/s. 
\end{abstract}

\maketitle

\section{Introduction}
The solar f modes are useful tools for studying perturbations that reside
within the top $2$~Mm below the surface of the Sun, and have been used as 
diagnostics as such before \citep{duvall2000}. The linear forward problem of
time-distance helioseismology is to calculate functions, or kernels, which
give the sensitivity of travel-time measurements to small-amplitude perturbations in the Sun. For a discussion of this general procedure,
which is the one that we will follow  here, see \citet{gizon2002}. 

In this study we are interested in flow kernels, i.e. functions
which give the sensitivity of travel-time measurements to steady, small amplitude, two-dimensional, spatially-varying horizontal flows. By flow kernel, we mean the 
function $\bK$ defined by
\begin{equation}
\tau_{\rm diff}(\1|\2)=\int\!\!\!\!\int \bu(\bx)\cdot\bK(\bDelta,\bx) \,  \id^2\bx  ,
\label{kern-def}
\end{equation}
where $\tau_{\rm diff}$ is the difference in travel time for waves propagating in opposite directions between two fixed locations $\1$ and $\2$ on the surface, $\bu=(u_x,u_y)$ is the two-dimensional flow vector, and the integral is over the position vector $\bx=(x,y)$ that spans the whole solar surface. We adopt a cartesian geometry for the sake of simplicity. The kernel $\bK=(K_x,K_y)$ is a two-dimensional vector which gives the sensitivity to flows in the $\hat{\bx}$ and $\hat{\by}$ directions. Furthermore, we define $\bDelta=\2-\1$ and the distance $\Delta= \|\bDelta\|$ between the two points $\1$ and $\2$.

In Section~\ref{sec:observed} we describe our simplified model of f-mode propagation through an inhomogeneous moving medium. In Section~\ref{sec:uniform}, we consider a uniform flow and compare the exact solution, its linearization in the flow, and the third-order approximation. In Section~\ref{sec:kernels}, we show example kernels. We discuss the results in the last section.


\section{Zero-order wavefield and first Born approximation}
\label{sec:observed}
In general, observations of solar oscillations are described by the filtered line-of-sight projection of the observed Doppler velocity
$\bbv$ given by
\begin{equation}
\label{psi-phi}
\psi(\bx, t) =\filter [\los\cdot\bbv ] \equiv\filter [ \phi ],
\end{equation}
where the operator $\filter$ describes the convolution with the point-spread function of the telescope and any additional filtering used in the data analysis, $\los$ denotes the line-of-sight unit vector, and $\phi$ is the line-of-sight Dopplergram.

As was done by \citet{gizon2002}, we consider a constant density half-space with a free surface at $z=0$. Gravitational acceleration, $-g\zhat$, is constant. The medium is permeated by a steady, horizontal, divergenceless, and irrotational flow, $\bu(\bx)$. The condition $\bnabla\cdot\bu=0$ is imposed by the continuity equation in the steady state. In addition, we require that $\bnabla\times\bu=0$ to keep the problem two-dimensional. Under these conditions of smoothness, the fluid velocity remains irrotational at all times. 

As is appropriate for solar waves, we consider small-amplitude (linear) waves. At the free surface $z =0$, the linearized dynamical and kinematic boundary conditions are
\begin{eqnarray}
p/\overline{\rho} -  g \eta &=& \Pi/\overline{\rho} \qquad z=0,\\
-\ii \omega \eta + \bu \cdot \bnabla \eta &=& \partial_z \Theta  \qquad z=0,
\end{eqnarray}
where $\Theta$ is the wave velocity potential $\bbv=\bnabla\Theta$, $\eta$ is the elevation of the fluid's surface, $p$ is a pressure perturbation, $\overline{\rho}$ is
the background density, and $\Pi$ is a stochastic pressure source that generates the waves. The above equations and the following ones apply to quantities that were Fourier-transformed in time ($\omega$ is the angular frequency). In the bulk, the linearized momentum and continuity equations are
\begin{eqnarray}
-\ii \omega \Theta + \bu\cdot\bnabla\Theta &=& - p/\overline{\rho} - \Gamma[\Theta],\\ 
\bnabla^2 \Theta &=& 0, 
\end{eqnarray}
where $\Gamma$ is a phenomenological frequency-dependent damping operator \citep{gizon2002}.

In order to obtain the scattered wavefield computed in the first Born approximation, we expand any wave quantity $q$ into a zero-order term $q^0$ (in the absence of the flow) and a first-order term $\delta q$ caused by the flow perturbation. In the bulk, the velocity potentials $\Theta^0$ and $\delta\Theta$ both satisfy the three dimensional Laplace's equation. Eliminating pressure and surface elevation, the zero-order and first-order surface boundary conditions at $z=0$ reduce to 
\begin{align}\label{source}
(\partial_z -\kappa)\Theta^0 &= \frac{\ii\omega}{\overline{\rho} g} \Pi\equiv S^0,\\
(\partial_z -\kappa)\delta\Theta &= \ii \bu\cdot\bnabla \left[(\partial_\omega\kappa)\Theta^0 + \partial_\omega S^0\right] \equiv \delta S  ,
\end{align}
where  $\kappa(\omega) = \omega^2/g  + \ii \gamma\omega/g$ is the complex
wavenumber at resonance and  $\gamma(\omega)$ is a realistic damping rate \citep{gizon2002}. In doing so, we have used the approximations $\delta\Pi=\ii \bu\cdot\bnabla \partial_\omega\Pi$ and $\delta \Gamma \Theta^0 = \ii  (\partial_\omega\gamma) \bu\cdot\bnabla  \Theta^0$, which describe the perturbations to the source of excitation and the wave damping respectively (this can be understood by transforming to a co-moving frame).

The solution of the zero- and first-order problems can be solved by introducing
a Green's function. The final result for the observable $\psi$ (on the surface) is 
\begin{align}
\label{psi0}
\psi^0 (\bx, \omega) &=  \filter\left[ (\ell_{\rm h}\cdot\bnabla_{\rm h} + \ell_z\kappa) \Theta^0   \right], \\\label{dpsi}
\delta\psi(\bx,\omega)  &= \filter\left[ (\ell_{\rm h}\cdot\bnabla_{\rm h} + \ell_z\kappa) \delta \Theta  + \ii \ell_z(\partial_\omega\kappa) \bu\cdot\bnabla \Theta^0   \right],
\end{align}
where $\los=(\ell_{\rm h}, \ell_z)$, $\bnabla_{\rm h}$ is the horizontal gradient, and the zero-order and first-order surface velocity potentials are given by
\begin{align}
\Theta^0(\bx,\omega)  &=  2\pi\int\!\!\!\int  G(\bx-\bx',\omega)S^0(\bx',\omega) \id^2\bx' ,  \\
\delta\Theta(\bx,\omega) & =  2\pi \int\!\!\!\int  G(\bx-\bx',\omega) \delta S(\bx',\omega) \id^2\bx' ,
\end{align}
with 
\begin{equation}
G(\bx,\omega) = \frac{1}{(2\pi)^3} \int\!\!\!\int \frac{e^{\ii\sbk\cdot\sbx}}{k-\kappa(\omega)} \, \id^2\bk ,
\end{equation}
where $k = \| \bk\|$ is the horizontal wavenumber. With the above expressions for the observed wavefield, one can compute the temporal cross-covariance function.


\section{Uniform flow: exact, first-, and third-order solutions}
\label{sec:uniform}

\begin{figure*}[t]
\begin{minipage}[t]{0.48\textwidth}
\centering
\includegraphics[width=\textwidth]{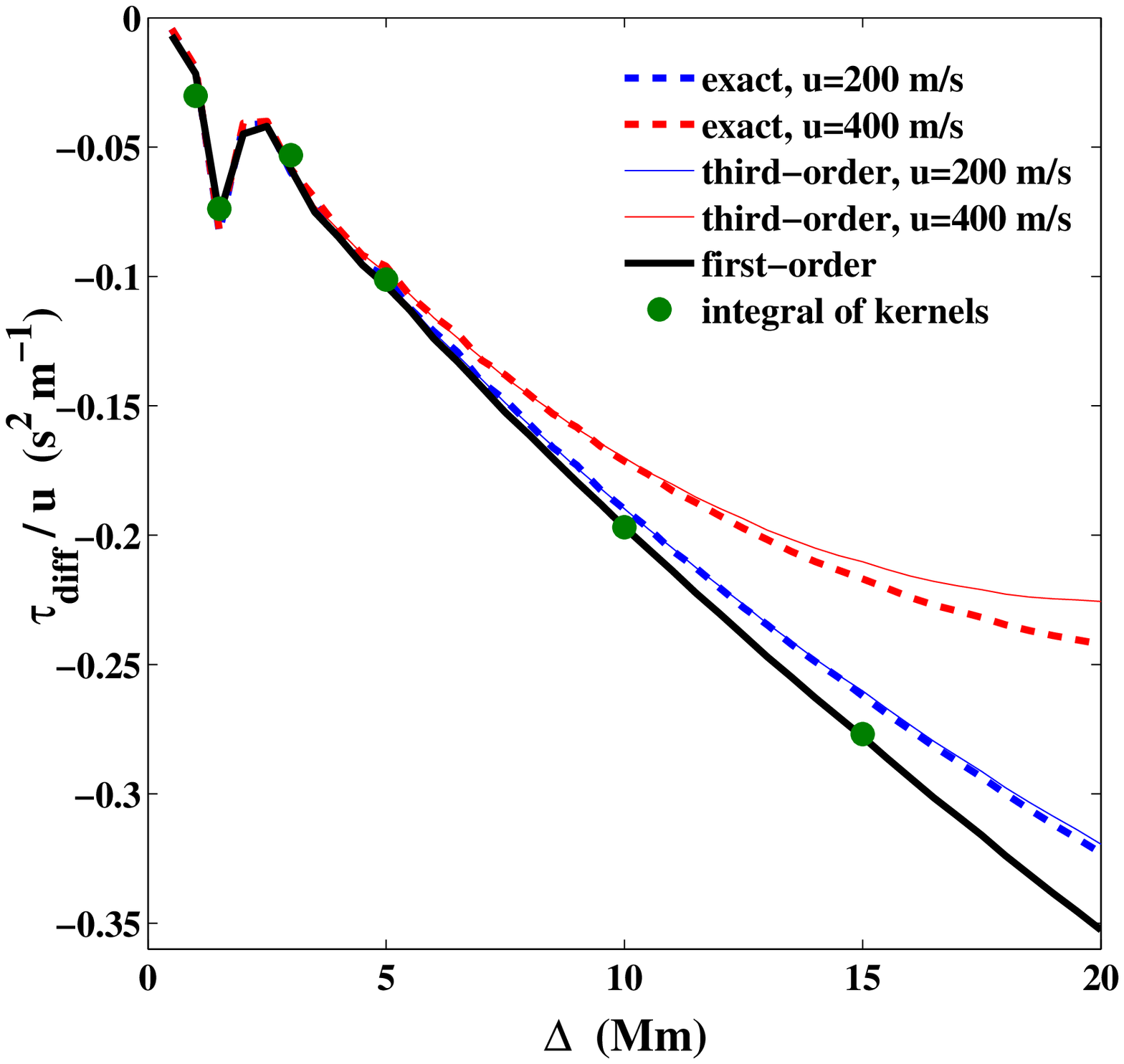}
\caption{F-mode travel-time differences versus distance for uniform horizontal flows. Two values are chosen for the flow amplitude, $u=200$~{\rm m/s} and $u=400$~{\rm m/s}. The thick black solid line is the solution to the first-order
  approximation, the thin solid lines are the third-order solutions, and the dashed lines are the exact solutions. The green dots denote the values of the
  spatially-integrated kernels from Sec.~\ref{sec:kernels}.}
\label{fig:tt_v_d}
\end{minipage} 
\hfill
\begin{minipage}[t]{0.48\textwidth}
\centering
\includegraphics[width=\textwidth ]{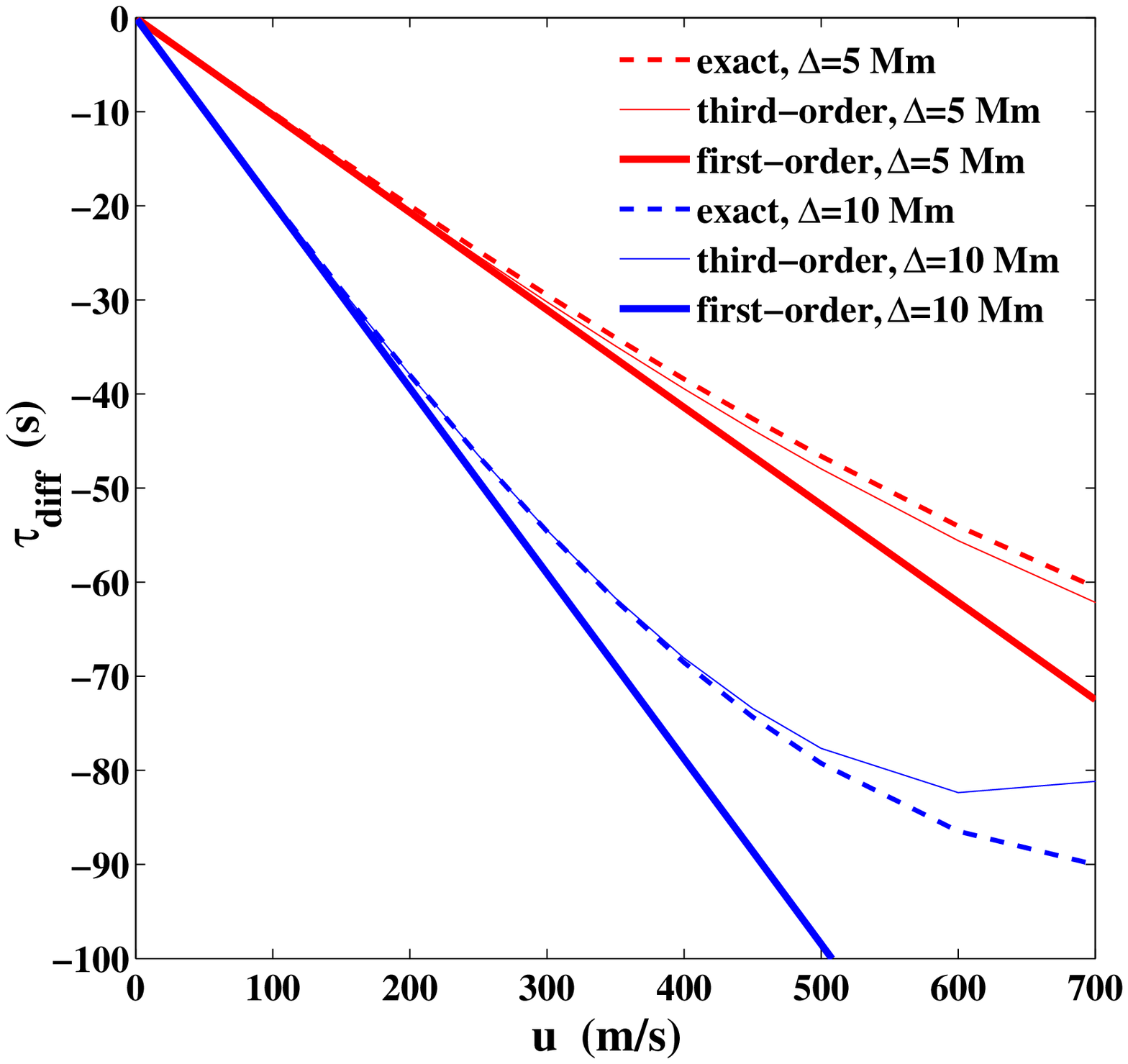}
\caption{F-mode travel-time differences versus uniform horizontal flow for different distances. Two distances are studied, $\Delta=5$~{\rm Mm} and $\Delta=10$~{\rm Mm}, given by the red and blue sets of curves respectively. For each set of curves, the thick solid line denotes the first-order solution, the thin solid line is the third-order solution, and the dashed line is the exact solution.}
\label{fig:tt_v_u}
\end{minipage}
\end{figure*}

Before using the first-order Born approximation from the previous section to derive the sensitivity kernels (see Sec.~\ref{sec:kernels}), we wish to consider here the case of constant flows,
\begin{equation}
 \bu = {\rm const}.
\end{equation}
 In this case,  $\phi(\bx,t)$ can be calculated simply by considering the Galilean transformation $\bx \rightarrow \bx-\bu t$:
\begin{equation}
\phi(\bx,t)=\phi^0(\bx-\bu t,t) .
\end{equation}
This transformation applies to all other quantitites in the problem except for the filter $\filter$. So the filtered wavefield in Fourier space is then given by
\begin{equation}
\label{psi-flow} 
\psi(\bk,\omega) = F(\bk,\omega) \phi^0(\bk,\omega-\bk\cdot\bu),
\end{equation}
where multiplication by $F(\bk,\omega)$ accounts for the action of the filter operator $\filter$ in Fourier space. The filter contains a mode-mass correction which is the ratio of the mode inertia in our model to the mode inertia in a standard stratified model. In general it also contains a phase-speed filter, but in the particular case presented here where only f modes are excited, a phase-speed filter is unnecessary and therefore $F$ is independent of $\omega$.

The expectation value of the cross-covariance function is given by \citep{gizon2002}
\begin{equation}
\label{c}
C(\1|\2,\omega)= \int\!\!\!\int
P(\bk,\omega) e^{\ii\sbk\cdot\sbDelta} \, \id^2\bk ,
\end{equation}
where 
\begin{equation}
\label{pow}
P(\bk,\omega) = E[|\psi(\bk,\omega)|^2]
\end{equation}
denotes the expectation value of the power spectrum. In particular, the zero-order, or unperturbed ($\bu=0$), cross-covariance is given by
\begin{equation}
\label{c0}
C^0 (\1|\2,\omega)= \int\!\!\!\int 
P^0(\bk,\omega) e^{\ii\sbk\cdot\sbDelta} \, \id^2\bk .
\end{equation}

Following \citet{gizon2002},  the travel-time difference can be measured from the cross-covariance function according to
\begin{equation}
\label{tau-w}
\tau_{\rm diff}(\1|\2)=\frac{-2{\rm Re}\int_0^{\infty} \ii\omega C^0(\bDelta,\omega)\,\Delta C(\bDelta,\omega) \, \id\omega  }{\int_0^{\infty}\omega'^2 |C^0(\bDelta,\omega')|^2 \, \id\omega'},
\end{equation}
using $\Delta C=C-C^0$, where Re takes the real part of the expression. The travel-time shift $\tau_{\rm diff}(\1|\2)$, i.e., the difference in time it takes for waves going from $\1$ to $\2$ and for those going from $\2$ to $\1$. In our problem, $\tau_{{\rm diff}}$ is due to the flow.

On one hand, the travel-time difference may be computed exactly for any value of $\bu$. This is simply done numerically by using eqs.~(\ref{psi-flow})-(\ref{tau-w}).

On the other hand, an approximation to the travel-time difference  can be obtained by expanding the power spectrum as a Taylor series in $\bu$.  This will provide us with a means to test the first-order Born approximation developed in Section~\ref{sec:observed}, as well as to quantify higher-order terms.  For a constant flow, the Taylor expansion of the power spectrum (eq.~[\ref{pow}]) is
\begin{align}
\label{expansion}
\nonumber
P (\bk,\omega) = &  P^0 (\bk,\omega) -\bk\cdot\bu |F|^2\partial_{\omega}|\phi^0|^2 \\ \nonumber
&+
\frac{1}{2}(\bk\cdot\bu)^2|F|^2\partial_{\omega}^2|\phi^0|^2\\
&
-\frac{1}{6}(\bk\cdot\bu)^3|F|^2\partial_{\omega}^3|\phi^0|^2 +\dots
\end{align}
We may truncate this expansion after each successive term to study the first-, second-, and third-order travel-time differences (using eq.~[\ref{tau-w}]). The second-order term does not affect travel-time differences because it only introduces perturbations to the cross-covariance that are symmetric in the time lag.

Figure~\ref{fig:tt_v_d} shows the exact and approximate travel-time differences $\tau_{\rm diff} / u$ versus $\Delta$ computed at fixed $\bu = u \hat{\bDelta}$ for $u=200$~m/s and $u=400$~m/s. We find that the first-order approximation is within 10\% of the exact value for $\Delta<20$~Mm, while the error is only about 1\% when terms up to $u^3$ are kept.  For a flow that is twice as large, $u=400$~m/s, non-linear effects are more important. The non-linearity of $\tau_{\rm diff}$ with $u$ also increases as the separation distance $\Delta$ increases.

Figure~\ref{fig:tt_v_u} shows the variations of $\tau_{\rm diff}$ as a function of $u$ at fixed distances $\Delta=5$~Mm and $\Delta=10$~Mm. The first-order approximation is reasonable for flows with amplitudes less than about 400~m/s, although it gets worse for larger distances.  In particular this means that the first-order approximation (and thus the kernels given in Sec.~\ref{sec:kernels}) is appropriate to study supergranular flows.


\section{Sensitivity Kernels}
\label{sec:kernels}

\begin{figure*}[t]
   \includegraphics[ width=7 in]{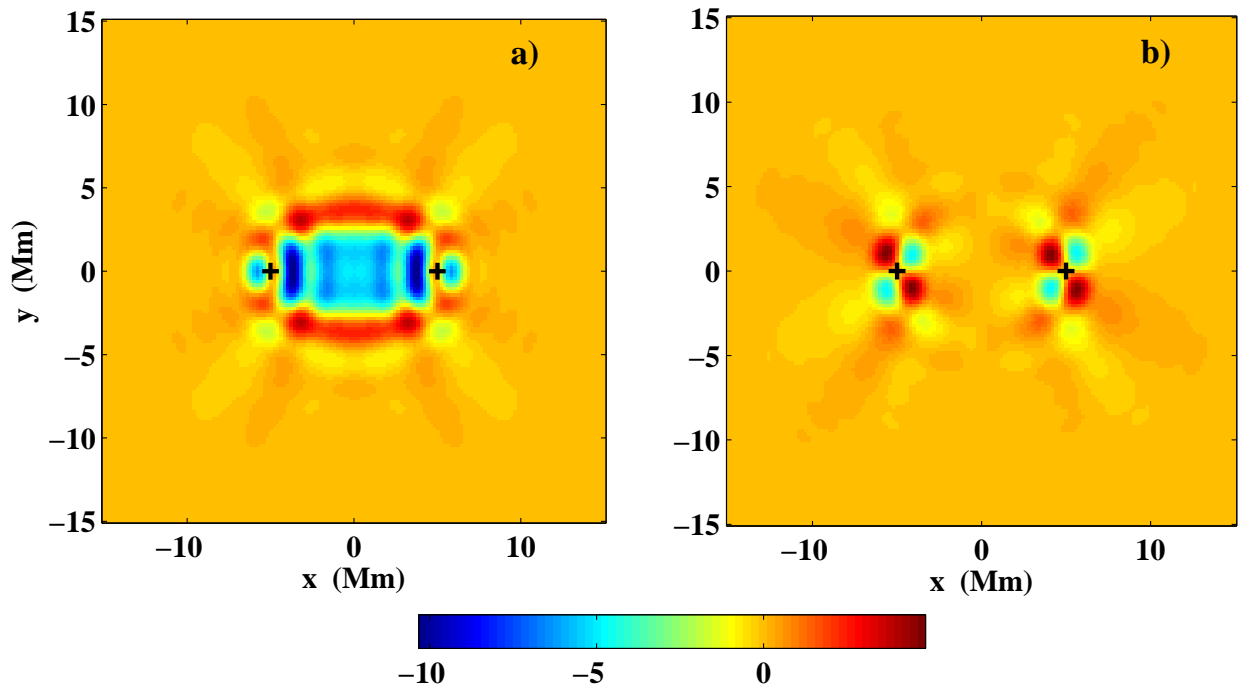}
\centering
    \caption{F-mode travel-time sensitivity kernels for flows. The separation distance of the observation points, denoted by the crosses, is $\Delta=10$~{\rm Mm}. The coordinates are $\1=(-5,0)$~{\rm Mm} and $\2=(5,0)$~{\rm Mm}. The units of the colorscale are {\rm s\,Mm$^{-2}$/(km/s)}. The $K_x$ and $K_y$ components of the kernel are shown in panel {\rm a)} and panel {\rm b)} respectively.}
  \label{fig:kerns10}
\end{figure*}

We now return to the computation of sensitivity kernels in the first Born approximation in order to connect $\tau_{\rm diff}$ with a spatially varying flow $\bu(\bx)$.  

The first-order perturbation to the travel time is obtained by approximating $\Delta C$ in eq.~(\ref{tau-w}) by 
\begin{align}
\label{deltac}
\nonumber
\Delta C(\1,\2,\omega)\simeq & E\left[\delta\psi^*(\1,\omega)\psi^0(\2,\omega)\right.\\
&+\left.\psi^{0*}(\1,\omega)\delta\psi(\2,\omega)\right]. 
\end{align}
The linear dependence of $\Delta C$ and $\tau_{\rm diff}$ on $\bu$ then follows from the expression of $\delta\psi$ (eq.~[\ref{dpsi}]).

The kernel functions are identified according to their definition, eq.~(\ref{kern-def}).  The explicit expression of the flow kernels will be given in an
upcoming publication.

A pair of sensitivity kernels is given in Fig.~\ref{fig:kerns10} for a distance
$\Delta=10$~Mm (the observation points have coordinates $\1=(5,0)$~Mm and $\2=(-5,0)$~Mm). The kernel on the left, $K_x$, gives the sensitivity to $u_x$ and the
kernel on the right, $K_y$, gives the sensitivity to $u_y$. The flow kernel $K_x$ displays elliptical and hyperbolic features, just as the kernels for the damping rate and the source strength derived earlier by \citet{gizon2002}. The elliptical features have been called Fresnel zones in geophysics (e.g., \citep{Tong}). The first Fresnel zone has a size (measured along $x=0$) which is roughly given by $(\lambda\Delta)^{1/2}$, where $\lambda\simeq 5$~Mm is the dominant f-mode wavelength. The hyperbolae are due to scattering of waves generated by distant excitation events. For reasons of symmetry (the $K_y$ kernel is antisymmetric with respect to the lines $x=0$ and $y=0$) the total integral of $K_y$ is zero.


For consistency, we have checked that the total integral of $K_x$ is nearly exactly the same (within 0.1\%) as the first-order approximation of $\tau_{\rm diff}/u$ calculated under the assumption of a constant flow (Sec.~\ref{sec:uniform} and Fig.~\ref{fig:tt_v_d}). This gives us confidence in the numerics.

The effect of the line-of-sight projection of the Doppler velocity signal on the flow kernels are included in the model of Sec.~\ref{sec:observed}. This is not presented here, although it is briefly discussed in \citet{Jack2006}.


\section{conclusions}

Sensitivity kernels for flows have been calculated in the first Born approximation following the general recipe given by \citet{gizon2002}.  

It has proven very useful to consider a uniform flow model, which allows one to find an exact solution for the travel times, as well as approximations to the exact solution in leading powers of the flow $u$. 

It was shown that the first-order solution is equivalent to the first Born approximation in the case of a uniform flow, and thus provides a test of the consistency of the kernels. This test revealed an accuracy within about 0.1\%, giving us confidence in the numerical calculations. Furthermore, comparisons of the first-order solution with the exact solution show that it is reasonable to study flows up to 400~m/s, with the caveat that this value depends on the distance $\Delta$, with larger distances giving more error. Nonetheless, the linear model presented here is likely quite capable for the study of supergranular flows.



\end{document}